\documentstyle[multicol,aps,psfig]{revtex}
\newcommand{\be}{\begin{equation}}
\newcommand{\ee}{\end{equation}}
\newcommand{\bea}{\begin{eqnarray}}
\newcommand{\eea}{\end{eqnarray}}
\begin{document}
\title{Time Delay Induced Death in Coupled Limit Cycle Oscillators}
\author{D.V. Ramana Reddy\cite{foot_1}, A. Sen and 
G.L. Johnston\cite{foot_2}}
\address{Institute for Plasma Research,\\ 
Bhat, Gandhinagar 382428, India.}
\date{\today}
\maketitle
\begin{abstract}
We investigate the dynamical behaviour of two limit cycle oscillators
that interact with each other via time delayed coupling and find that
time delay can lead to amplitude death of the oscillators even if they
have the same frequency. We demonstrate that this novel regime of
amplitude "death" also exists for large collections of coupled
identical oscillators and provide quantitative measures of this death
region in the parameter space of coupling strength and time delay.
Its implication for certain biological and physical
applications is also pointed out.

\end{abstract}
%
\begin{multicols}{2}
Coupled limit cycle oscillators provide a simple but powerful 
mathematical model for simulating the collective behaviour of a
wide variety of systems that are of interest in physics
\cite{AKJ_84,SM_88,AEK_90,YA_76,MS_90,MS_90a,YK_79,SSW_92,GBE_90,HD_90},
chemistry \cite{KBE_85,CE_89} and biological sciences
\cite{ATW_80,KS_80}.  These oscillators have also attracted some
large scale numerical \cite{KS_89} and novel experimental efforts
\cite{BL_96}.  For weakly coupled oscillators the predominant effect is
a synchronization of the frequencies of the individual oscillators to a
single common frequency once the coupling strength exceeds a certain
threshold, while the amplitudes remain unaffected. For stronger
couplings the amplitudes also play an important role and give rise to
interesting phenomena like the Bar-Eli effect \cite{KBE_85} where all
the oscillators suffer an amplitude quenching or {\it death}
\cite{MS_90a,GBE_90}. 
In general there can be a wide variety of collective behaviour including
partial synchronisation, phase trapping, large amplitude Hopf
oscillations and even chaotic behaviour\cite{AEK_90,YA_76,MS_90}.
In recent times there have been extensive investigations of 
coupled oscillator systems including elegant statistical 
mechanics formulations in the limit of infinite number of 
oscillators \cite{YK_79,HD_90}.

The salient features of the behaviour of a finitely large number
of oscillators (usually obtained from numerical or approximate 
analytic means) can often be understood by analysing just two
coupled oscillators. We have carried out such an analysis to
investigate the effect of time delay on the interaction between
two limit cycle oscillators. Time delay is ubiquitous in most
physical and biological systems \cite{GM_77,CG_86,HS_88},
arising from finite propagation
speeds of signals for example, and have not been widely studied
in the context of coupled limit cycle oscillator systems.
Niebuhr {\it et al} \cite{NSK_91} and Schuster and Wagner \cite{SW_89}
who are one of the few
who have carried out such an investigation,  have restricted
themselves to the simpler {\it coupled phase}  models where the
phenomenon of amplitude death does not exist.  In our model
equations we have retained both the phase and amplitude response
of the oscillators and we find that time delay has a significant
effect on the characteristics of all the major cooperative
phenomena like frequency locking, phase drift and amplitude
deaths. In particular our detailed numerical investigations show
that in the presence of time delay the parameter regime of
amplitude death can extend down to the region of zero frequency
mismatch between the oscillators.  This is in sharp contrast to
the situation with no time delay where all previous numerical
and analytical studies \cite{AEK_90,MS_90,MS_90a,GBE_90}
show that amplitude death can occur only
if the coupling between oscillators is sufficiently strong and
when the frequencies are sufficiently disparate.
In this Letter
we confine ourselves primarily to the effect of time delayed
coupling on the phenomenon of amplitude death and present a
detailed numerical and analytical estimate of the parameter
space in coupling strength and time delay where such a death can
occur for identical oscillators.  We also establish that this
effect is not an artefact of the simple two oscillator model,
but can occur for a system of large number of globally (or
locally) coupled
identical oscillators (including the continuum limit of $N
\rightarrow \infty$).

We analyse the following model equations:
\bea
\label{z1}
\dot{Z}_1(t)=(1+i\omega _1-\mid Z_{1}(t)\mid ^2)Z_1(t) \nonumber \\
+K[Z_2(t-\tau )-Z_1(t)],\\
\label{z2}
\dot{Z}_2(t)=(1+i\omega _2-\mid Z_{2}(t)\mid ^2)Z_2(t) \nonumber \\
 +K[Z_1(t-\tau )-Z_2(t)],
\eea
\noindent
where $\tau $ is a measure of the time delay, $K$ is the coupling
strength, $\omega _{1,2} $ are the intrinsic frequencies of the
two oscillators and $Z_{1,2}$ are complex. The model is a 
generalization of the diffusively and linearly coupled oscillators
studied extensively for example in \cite{AEK_90,KBE_85}. The
time delay parameter is introduced in the argument of the
coupling oscillator (e.g. $Z_{2}$ in (1)) to physically account
for the fact that its phase and amplitude information is
received by oscillator $Z_{1}$ only after a finite time $\tau$
(due to finite propagation speed effects).  In the absence
of coupling ($K=0$) each oscillator has a stable limit cycle at
$\mid Z_{i}\mid =1$ on which it moves at its natural frequency
$\omega_{i}$.  The coupled equations represent the interaction
between two weakly nonlinear oscillators (that are near a Hopf
bifurcation) and whose coupling strength is comparable to the
attraction of the limit cycles.  It is important then to retain
both the phase and amplitude response of the
oscillators\cite{AEK_90}.  The state $Z_{i}=0$ is an equilibrium
solution of the system of equations [(\ref{z1}) and (\ref{z2})]. For
$K=0$ this equilibrium state is linearly unstable since the
individual oscillators tend to stable limit cycle states $\mid
Z_{i}\mid =1$. The stability of {\it amplitude death} for $K
\neq 0$ has been studied in great detail by Aronson {\it et
al}\cite{AEK_90} for system ((\ref{z1},(\ref{z2})) in the absence
of any time delay in the coupling (i.e. for $\tau=0$). The
conditions for stability found by them are,
\be
\label{aron}
K > 1 \;\;\; and \;\;\; \Delta = \mid \omega_{1} - \omega_{2} \mid
 > 2\sqrt{2 K -1},
\ee 
which shows that amplitude death can occur in this case only for
sufficiently large values of $\Delta$ provided $K>1$.

In Fig. 1(a), we reproduce the bifurcation diagram of 
Aronson {\it et.al}\cite{AEK_90}
where the region marked $I$ represents the amplitude death region and
the dotted curves mark the boundary as defined by condition
(\ref{aron}). The two bounding curves intersect at the point
($K=1, \Delta =2$).  Regions marked ($II$) and ($III$) represent
phase locked and phase drift regions respectively which we will
not discuss in detail here. In Fig. 1(b) we present the bifurcation 
diagram of (\ref{z1}-\ref{z2}) for $\tau=0.0817$. 
Note that in contrast to the diagram of Fig. 1(a), 
the amplitude death region now extends down to
$\Delta = 0$ and has a finite extent along the coupling strength
($K$) axis. We find that the phenomenon persists for a range of
$\tau$ after which the bifurcation curve lifts up from the
$\Delta = 0$ line and identical oscillators can no longer suffer death.  
Fig. 2(a) shows this region, for different values of $\omega$,
in $\tau - K$ space for which amplitude
death of identical oscillators
can occur.  The size of this {\it death island} is a function of the
frequency of the oscillators ($\omega$), as shown by the other
curves. We shall soon show that the size is also a function
of $N$, the number of oscillators.  The bifurcation curves (including
the island boundaries) have been obtained from a linear stability
analysis of (\ref{z1}),(\ref{z2}) about the origin $(Z_{1} = Z_{2}
=0$) as well as direct numerical integration of the equations.
Assuming the linear perturbations to vary as $e^{\lambda t}$ the 
\begin{figure}
\narrowtext
\centerline{\hbox{\psfig{file=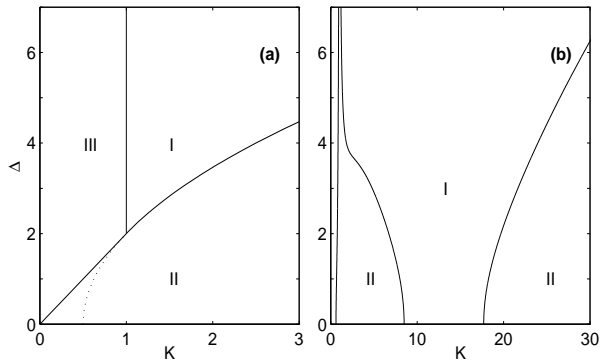,width=8cm,height=5cm}}}
\caption{Bifurcation diagram of Eqns.(\ref{z1}-\ref{z2}).
(a) $\tau=0$. Region $I$ is the amplitude death
region, $II$ is the phase locked region, and $III$ corresponds to the
phase drift or incoherent region. (b) $\tau =
0.0817$, and $\bar\omega = 10$. The death region extends down to
$\Delta = 0$ indicating that identical oscillators can suffer amplitude
death. The phase locked region is split into two disjoint regions.}
\end{figure}
\noindent
characteristic eigenvalue equation we get is,
\bea
\label{char}
(1-K + i\omega_{1} - \lambda )( 1-K + i\omega_{2} - \lambda) - \nonumber \\
K^{2} e^{-2\lambda \tau} = 0,
\eea
where $\lambda$ is the complex eigenvalue and the complete set of
eigenvalues includes those arising from the complex conjugate
equations of (1) and (2) . 
Setting 
${\it Real}(\lambda) = 0$ in (\ref{char}) and separating
the real and imaginary parts, the equations for the
critical curves (i.e. the marginal stability condition) 
are,
\bea
\label{real}
\lambda_{I}^{2} + 2\lambda_{I} \bar{\omega} + \bar{\omega}^{2}
- \frac{\Delta^{2}}{4} - (1-K)^{2} \nonumber  \\
+ K^{2}\cos(2\lambda_{I}\tau) = 0, \\
\label{imag}
2(1-K)(\bar{\omega} + \lambda_{I}) + K^{2} \sin(2\lambda_{I}\tau) = 0.
\eea
where $\lambda_{I} = {\it Imag}(\lambda)$ and $\bar{\omega} =(\omega_{1}
+\omega_{2})/2$ is the mean frequency. Eliminating $\lambda_{I}$
between  (\ref{real}) and (\ref{imag}) and considering the full
set of eigenvalues , we obtain the following
transcedental relation between $K, \Delta$ and $\tau$
which is now the modified marginal stability condition
in place of (\ref{aron}),
\be
\label{marg}
g \alpha = K^2 \sin (\alpha \tau \pm 2\bar{\omega}\tau),
\ee
where $\alpha =\sqrt{\Delta ^2-4g^2\mp 4\sqrt{K^4-g^2\Delta ^2}}$, and
$g=1-K$. Note that for $\tau=0$, the above relation readily
simplifies to (\ref{aron}) and yields the marginal stability
curves $K=1$ and $2K=1+\frac{\Delta ^2}4$. Figure 1(b). is 
a numerical plot of (\ref{marg}) for $\tau = 0.0817$.

To obtain a condition for the death of identical
oscillators we repeat the analysis with $\omega_{1} = \omega_{2}
=\omega$ in (\ref{char}) and after eliminating 
$\lambda_{I}$, obtain the relations
\be
\label{tauk1}
\tau = \frac{\cos^{-1}(1-1/K)}{\omega - \sqrt{2K-1}}\;;\;\;
\tau = \frac{\pi - \cos^{-1}(1-1/K)}{\omega + \sqrt{2K-1}}.
\ee
\begin{figure}
\narrowtext
\centerline{\hbox{\psfig{file=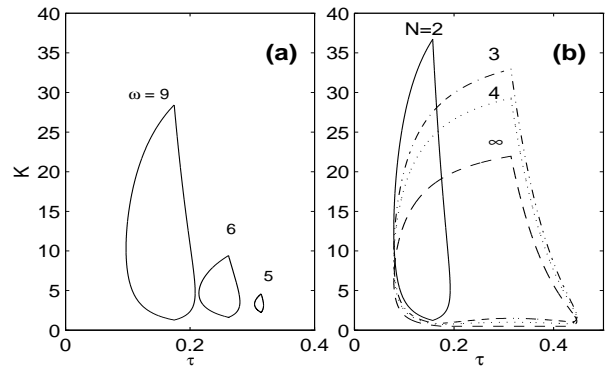,width=8cm,height=5cm}}}
\caption{(a) The region of amplitude death for $N=2$ as a function of
the common intrinsic frequency. The size of the the island decreases
with decreasing frequency and vanishes below a certain threshold.
(b) The death islands as a function of the number of globally coupled
oscillators.  Each oscillator is assumed to have an intrinsic frequency
of $\omega= 10$.  The death island survives even in the limit of $N
\rightarrow \infty$.}
\end{figure}
The region of intersection between the two curves (\ref{tauk1})
corresponds to the {\it death island} region of Fig. 2(a).
It demonstrates the only stability switches
(of the origin) that take place as a function of $\tau$.
For a given value of $\omega$ and
at a fixed $K$ we move (by varying $\tau$) from an unstable
region into a stable region as we cross the left boundary of the
island to emerge again into an unstable region as we cross the
right boundary of the island. No other {\it amplitude death}
islands are seen for larger values of $\tau$ (for a fixed $\omega$). 
We have also
confirmed this analytically by looking at the behaviour of the
supremum of the real parts of the roots of the transcedental
equation,
$d\lambda_{R}/d\tau$ (obtained from (4) with $\Delta = 0$) as
a function of $\tau$ in
the various parameter regimes\cite{CG_86}. The detailed mathematical proof
of this result will be published elsewhere.

Can this 
phenomenon occur for an arbitrary number of oscillators? To answer
this question we have investigated the following generalized set
of globally coupled equations:
\begin{eqnarray}
\label{nosc}
\dot{Z}_{i}(t)=(1+i\omega _{i}-\mid Z_{i}(t)\mid ^2)Z_{i}(t)+ \nonumber \\
\frac{2K}{N}\sum_{j=1}^{N}[Z_{j}(t-\tau)-Z_{i}(t)] - \nonumber \\
\frac{2K}{N}[Z_{i}(t-\tau) - Z_{i}(t)],
\end{eqnarray}
\noindent
where $i = 1,....,N$ and the last term on the right hand side has been included
to remove the self-coupling term.  For $\tau=0$, (\ref{nosc}) reduces
to the set of equations that have been extensively studied by
Ermentrout \cite{GBE_90}, Mirollo and Strogatz \cite{MS_90a}, and
others\cite{YA_76}.  Mirollo and Strogatz\cite{MS_90a} have
provided rigorous analytical and numerical conditions for amplitude
death in such a system. Their conclusions, in general, are similar to
the case of $N=2$, namely, that one needs a sufficiently large variance
in frequencies for death to occur and $K$ has to be sufficiently large.
We have been able to carry out a similar linear stability analysis of
(\ref{nosc}) around the origin for the case of finite $\tau$ and for a
large number of identical oscillators ($\omega_{j} = \omega,\;\;\; j=1,...,N$).
The resulting stability condition yields the following bounding
curves for the death island region:
\begin{eqnarray}
\tau = \frac{\cos^{-1}p}{\omega - \sqrt{4Kb-1}}\;;\;\;
\tau = \frac{2 \pi - \cos^{-1}p}{\omega + \sqrt{4Kb-1}}, \nonumber \\
\tau = \frac{\cos^{-1}(\frac{b}{b-1}p)}{\omega + \sqrt{q}}\;;\;\;
\tau = \frac{2\pi - \cos^{-1}(\frac{b}{b-1}p)}{\omega - \sqrt{q}},\nonumber \\
\end{eqnarray}
where $p = 1-1/(2Kb)$, $q= 4K^{2}-1+4Kb(1-2K)$ and the factor $b = (1-1/N)$ 
introduces the $N$ dependence of the island size explicitly. In
Fig. 2(b) we have plotted these islands for $N=2,3,4$ and
$N=\infty$.  
\noindent
To confirm these results we have also numerically
scanned the region with a direct numerical integration of
(\ref{nosc}) for a large number of oscillators, upto $N=10 000$,
and found excellent agreement. We have also carried out a
similar study for a large number of locally coupled identical
oscillators (nearest neighbour coupling \cite{SM_88}) with
periodic boundaries and find that time delay introduces death
islands in such systems as well.  Thus it appears that in the
presence of finite time delay in the mutual coupling, amplitude
death of identical oscillators is a fairly universal phenomenon
and occurs for any arbitrary number $N$ of oscillators extending
upto $N=\infty$ over a range of $\tau$ and $K$ values.  To the
best of our knowledge such a result has not been realised in the
past and may have important applications in biological or
physical systems. There are many physical examples of amplitude
death in real systems.  One of the earliest that was
investigated both theoretically and experimentally is that of
coupled chemical oscillator systems e.g. coupled
Belousov-Zhabotinskii reactions carried out in coupled stirred
tank reactors\cite{KBE_85,CE_89}. They can also occur in
ecological contexts where one can imagine two sites each having
the same predator-prey mechanism which causes the number density
of the species to oscillate. If the species are capable of
moving from site to site at a proper rate (appropriate coupling
strength) the two sites may become stable (stop oscillating) and
acquire constant populations. Another important application of
this concept is in pathologies of biological oscillator networks
e.g. an assembly of cardiac pacemaker cells\cite{ATW_80}.
Amplitude death signifies cessation of rhythmicity in such a
system which is otherwise normally spontaneously rhythmic for
other choices of parameters.  
For the onset of such an
arrhythmia, current models based on coupled oscillator networks
need to assume a significant spread in the natural frequencies
of the constituent cells (oscillators)\cite{MS_90a}. Our work
demonstrates that this assumption may not be necessary if one
takes into account time delay effects arising naturally from the
finite propagation times of the signals exchanged between the
cells. Another possible application is in the area of high power
microwave sources where it is proposed to enhance the microwave
power production by phase locking a large number of sources such
as relativistic magnetrons\cite{Ben_89}. Time delay effects,
arising from the finite propagation time of information signals
traveling through the connecting waveguide bridges, could impose
important limitations on the connector lengths and geometries in
these schemes. Our findings could provide a guideline in this
direction. It should be noted that a form of oscillator death
described in
\cite{EK_90} for identical oscillators is not a genuine amplitude death
since it occurs in the context of a {\it phase only} model. Time delay
in our study provides a new mechanism for genuine amplitude death to
occur in coupled identical oscillators.

Finally, it is worthwhile to mention that time delay can introduce
other interesting phenomena as well, some of which have been
studied in the context of the {\it phase only } model and need to
be investigated for the 
more general phase and amplitude model. Our numerical results, for
\begin{figure}
\narrowtext
\centerline{\hbox{\psfig{file=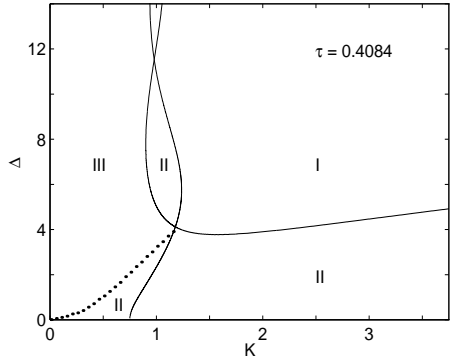,width=6cm,height=5cm}}}
\caption{The bifurcation diagram of Eqs. (\ref{z1}) and (\ref{z2}) for
$\tau = 0.4084$ and $\bar\omega = 10$. The amplitude death region (I)
is surrounded by the phase locked regions (II). The dotted curve
which separates the incoherent (III) and the phase locked regions is
obtained from numerical integration of the original equations.}
\end{figure}
\noindent
example, show that the bifurcation diagram of the system in
the presence of time delay has a significantly richer structure.
Fig. 3 is an example for the $N=2$ system
for $\tau = 0.4084$,  which can be contrasted with the
Aronson {\it et al}\cite{AEK_90} diagram of Fig. 1(a).
Note that one no longer has the clean separation of the Bar-Eli
region, the phase locked region and the phase drift region into
three disjoint regions that converge at a single degenerate
point. Instead the phase locked region now always surrounds the Bar-Eli
region and the single degenerate point is replaced by a
series of $X$ points resulting from the braided structure of the
phase locked region in the vertical direction. 
At large values of $K$
other bifurcation curves appear in the phase locked region
indicating the appearance of higher frequency states \cite{SW_89}. A detailed
investigation of various properties of this rich  phase diagram,
including stability studies of the various states, is now in
progress and will be reported elsewhere.

\end{multicols}

\end{document}